\documentclass[letterpaper,twocolumn,10pt]{article}
\usepackage{usenix}
\usepackage[available,functional]{usenixbadges}

\usepackage{amsmath,amsfonts}
\usepackage{graphicx}
\usepackage{booktabs}
\usepackage{array}
\usepackage{xcolor}
\usepackage{amsthm}
\usepackage{algorithm}
\usepackage{algorithmicx}
\usepackage[noend]{algpseudocode}
\usepackage{pifont}

\newcommand{\sysname}{{Magneton}}

\newif\ifdraft
\draftfalse

\newtheorem{hypothesis}{Hypothesis}

\newcommand{\cmark}{\ding{51}}
\newcommand{\xmark}{\ding{55}}

\begin{document}

\title{\LARGE{\textit{\sysname{}}: Optimizing Energy Efficiency of ML Systems via \\
Differential Energy Debugging}}

\author{
{\rm Yi Pan$^{1,3,*}$, Wenbo Qian$^{2,*}$, Dedong Xie$^1$, Ruiyan Hu$^{1,2}$, Yigong Hu$^2$, Baris Kasikci$^1$}\vspace{0.2em}\\
$^1$University of Washington, $^2$Boston University, $^3$UC Berkeley
}

\maketitle
\thispagestyle{empty}
{\let\thefootnote\relax\footnote{{$^*$Equal contribution. Correspondence to: Yi Pan <yi.pan@berkeley.edu>.}}}
\begin{abstract}
The training and deployment of machine learning (ML) models have become extremely energy-intensive. 
While existing optimization efforts focus primarily on hardware energy efficiency,
a significant but overlooked source of inefficiency is software energy waste caused by poor software design.
This often includes redundant or poorly designed operations that consume more energy without improving performance.
These inefficiencies arise in widely used ML frameworks and applications, yet developers often lack the visibility and tools to detect and diagnose them.

We propose differential energy debugging, a novel approach that leverages the observation that prominent ML systems often implement similar functionality with vastly different energy consumption.
Building on this insight, we design and implement \sysname{}$^\dagger$, an energy profiler that compares energy consumption between similar ML systems at the operator level and automatically pinpoints code regions and configuration choices responsible for excessive energy use.
Applied to 9 popular ML systems spanning LLM inference, general ML frameworks, and image generation, \sysname{} detects and diagnoses 16 known cases of software energy inefficiency and further discovers 8 previously unknown cases, 7 of which have been confirmed by developers.

{\let\thefootnote\relax\footnote{{$^\dagger$The code is available at \url{https://github.com/uw-syfi/Magneton}.}}}

\end{abstract}

\section{Introduction}

With the growing prevalence of large language models (LLMs)
and other machine learning (ML) applications,
the training and deployment of ML models have become massive, energy-intensive operations, often spanning thousands of GPUs in distributed clusters.
For example, generating a single response from Llama3-405B consumes over 3300 Joules (enough to power a 60W lamp for a minute) on an NVIDIA H100 GPU~\cite{chung2025mlenergybenchmarkautomatedinference}, while training GPT-4 is estimated to consume 50 GWh~\cite{economist2024energy}.
Optimizing ML system energy efficiency is therefore critical for both cost and sustainability.
In response, major companies and government agencies have launched initiatives to promote energy-efficient practices~\cite{google2022energy,huggingface2023energy,cra2025energywater}.

\begin{figure}[t]
    \centering
    \includegraphics[width=0.8\columnwidth]{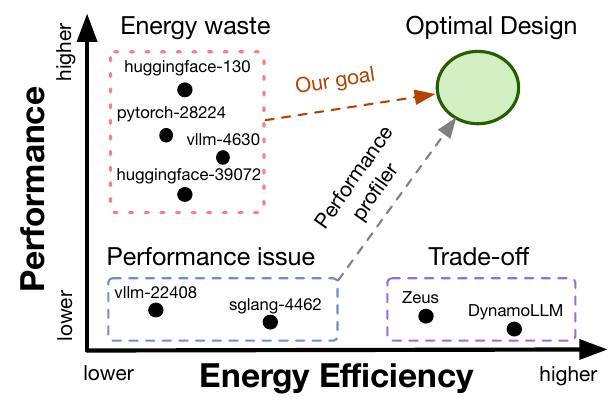}
    \vspace{-1em}
    \caption{Software energy waste, performance issues, and performance-energy trade-offs in the design space.}
    \vspace{-1.5em}
    \label{fig:case}
\end{figure}

Most existing energy optimization efforts focus on improving hardware energy efficiency, such as limiting GPU power, tuning GPU frequency, or scheduling tasks on low-power GPUs in heterogeneous clusters~\cite{Zeus2023NSDI,Perseus2024SOSP,DynamoLLM2025hpca,Splitwise2024ISCA}.
However, another significant but often overlooked source of energy inefficiency is poor software design, which we call \textit{software energy waste}.
For example, PyTorch’s DistributedDataParallel (DDP) manager, a package commonly used for data-parallel training, requires all GPUs to perform asynchronous all-reduce operations during data-parallel training. If data is imbalanced, GPUs that finish early still have to synchronize until the slowest worker completes its computation, rather than entering an idle state~\cite{pytorch-ddp}.
This issue is symptomatic of a broader class of problems in distributed ML systems, where subtle inefficiencies in communication or synchronization logic can lead to significant energy waste without impacting end-to-end latency.
Since hardware-centric methods cannot distinguish such redundant operations from useful work, they remain ineffective at addressing software energy waste.

Addressing software energy waste requires a two-pronged approach. First is detecting (identifying the symptom of) energy waste, and second is diagnosing (finding the root cause of) it. Automating the detection is important because modern ML systems often consist of millions of lines of code, and manually detecting software energy waste would be time-consuming and error-prone. Diagnosis is important, because even when an energy hotspot is found, developers may lack clear guidance on whether and how the energy hotspot can be restructured to drain less energy.

Ostensibly, performance profilers can help address software energy waste, but in practice they fall short. First, while some performance issues also lead to energy waste, \textit{software energy waste} (Figure~\ref{fig:case}) often causes little slowdown and is therefore difficult for performance profilers to detect. Furthermore, these tools find hotspots but provide little guidance on how to fix them. To make matters worse, existing energy profilers \cite{Batchsizer2021ASPDAC,Zeus2023NSDI} used in ML systems also fall short due to their coarse profiling granularity.
For instance, Zeus \cite{Zeus2023NSDI} requires a 100ms profiling window for accurate measurements---longer than the typical latency of an entire model.
This makes it impossible to detect waste at the granularity of individual operators, where it often occurs.
These profilers also fail to identify root causes or provide actionable guidance.

In this paper, we first study real-world ML systems and make two observations.
First, due to the widespread use of generative AI, each popular ML system---from distributed training systems like Megatron-LM and DeepSpeed to serving systems like vLLM and SGLang---typically has several competing alternatives with similar functionality.
Second, we analyze 8 widely-used ML systems~\cite{pytorch,jax,tensorflow,vllm,sglang,hf-transformers,stable-diffusion,diffusers} across three categories: LLM frameworks, ML libraries, and image generation and observe significant differences in energy consumption when performing the same tasks.
These findings suggest that the energy consumption differences among functionally similar ML systems can be leveraged to identify potential opportunities for reducing waste.

Building on this insight, we propose a new energy profiling method, \textbf{differential energy debugging}, to detect and diagnose software energy waste in ML systems. The basic idea is that when several ML systems implement the same task, their differences in implementation can lead to different energy footprints. By comparing these systems, we can detect energy-inefficient implementations whose energy consumption is significantly higher than their peers, and we can use the more efficient implementations to provide concrete guidance on how the wasteful implementation can be corrected.

A core challenge of differential energy debugging is that ML systems have very different internal code structures, which makes direct source-code comparison unreliable. 
This motivates the need for a common abstraction level at which different systems can be compared. A too coarse-grained abstraction level, such as a full transformer iteration, includes many operations together and therefore hides the root cause of energy waste. A too fine-grained abstraction level such as CUDA code is impractical due to unavailable source code and lack of hardware telemetry to measure energy at that level. 

We observe that ML systems rely on \textbf{operators} as their fundamental execution units. These operators, spanning both local computation (e.g., matrix multiplication) and distributed communication primitives (e.g., All-Reduce), provide a unified, semantically meaningful abstraction for comparison, as they construct common mathematical logic across diverse ML systems.
Moreover, they dominate the energy consumption of ML workloads, accounting for nearly 100\% of the total energy usage in our analysis of 8 ML systems.
Therefore, we argue that the operator is the natural abstraction level for differential analysis across different ML systems.

The second challenge is how to perform the comparison at the operator granularity to detect energy inefficiencies.
Naive approaches like one-to-one operator mapping or source-code analysis are infeasible, as many computational and data transfer tasks in ML systems are implemented by combinations of operators with large differences at the source code level.
To address this, we compare \textbf{the collection of operators} that perform the same task in the model, which we call semantically equivalent subgraphs.
By analyzing the dataflow in the computational graph, we match subgraphs with the same inputs and outputs across workloads, identifying them as semantically equivalent implementations of the same task.

The third challenge is diagnosing the source of energy waste. After detecting an energy hotspot, differential energy debugging needs to explain how and why different implementations of the same task diverge in their energy use. It analyzes the control flow and dataflow paths that lead from the semantically equivalent subgraphs back to their call sites. By comparing these paths across ML systems, it identifies the first point where the control flow diverges. This deviation is the likely root cause because it explains why the two systems invoke different combinations of operators and consume different amounts of energy.

To demonstrate the effectiveness of differential energy debugging,
we design and implement \sysname{}, an energy profiler for ML systems and 
apply it to 9 popular systems: vLLM, SGlang, Transformers, Megatron-LM, PyTorch, JAX, TensorFlow, Stable Diffusion, and Diffusers.
We collect 16 real-world cases of software energy waste, and \sysname{} detects and diagnoses 15 of them.
Furthermore, \sysname{} detects and diagnoses 8 previously unknown cases of software energy waste, among which 7 are confirmed by developers.
In summary, this paper makes the following contributions:
\begin{itemize}
    \item We propose differential energy debugging, a novel approach to identify and optimize software energy waste in ML systems.
    \item We design and implement \sysname{},
    an energy profiler that realizes differential energy debugging.
    \item We evaluate \sysname{} on 9 popular ML systems
    and demonstrate its effectiveness
    in detecting and diagnosing real-world software energy waste.
\end{itemize}

\section{Motivation}

\subsection{Case Studies}\label{sec:motivation_study}

To better understand the characteristics of software energy waste and why it is difficult to detect, we analyze three real-world cases collected from public GitHub issues and developer forums. These cases come from three widely used ML systems: PyTorch, an ML library; HuggingFace Transformers, an LLM inference framework; and Stable Diffusion, an image generation application.
These cases have different root causes and exhibit different energy waste patterns.
We selected these three cases to provide a representative coverage of a broad range of energy inefficiency issues in modern ML systems.

\begin{figure}[t]
    \centering
    \includegraphics[width=0.95\columnwidth]{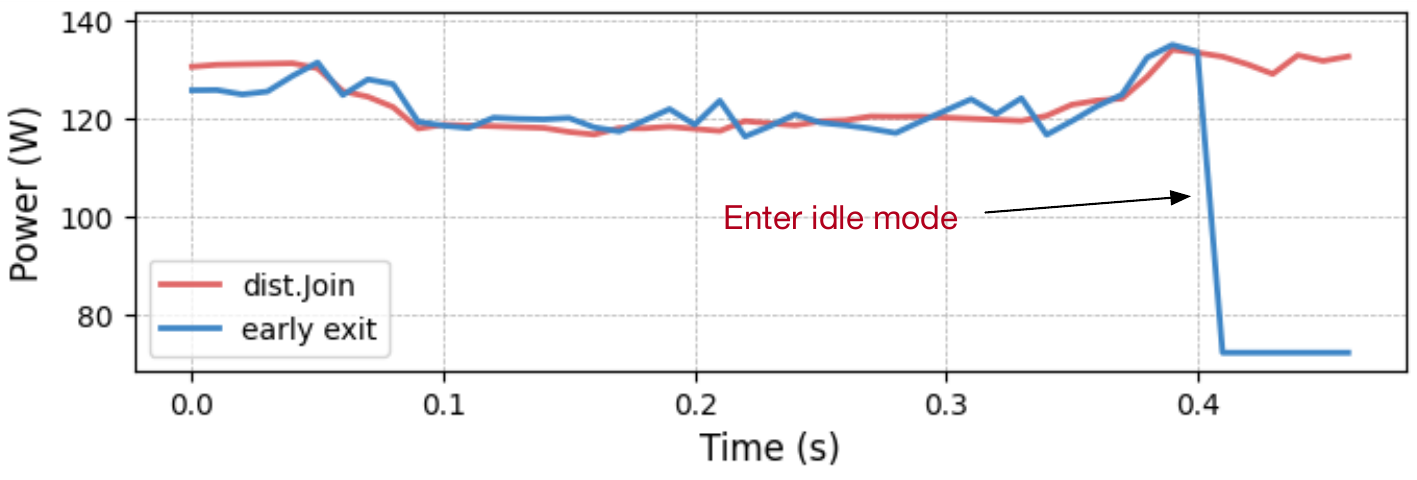}
    \caption{The power consumption of the join and early exit in DDP manager.}
    \label{fig:study_redundant_energy}
\end{figure}

\textbf{Case 1: Redundant Communication.}
In distributed training, the energy cost of a job is determined not only by the computation each worker performs
but also by the synchronization semantics that bind workers together: a
collective that requires participation from every rank forces stragglers and finishers alike to keep their GPUs resident.
PyTorch's DistributedDataParallel (DDP)~\cite{pytorch-DistributedDataParallel} is a widely used package for parallelizing model training across multiple GPUs.
It provides the \textit{Join} API for handling uneven inputs across GPUs, which forces finished GPUs to continue issuing all-reduce calls with zero-valued tensors so that the remaining GPUs' collectives can complete. However, these zero-tensor all-reduces are real NCCL kernels that occupy SMs on the GPU. As a result, the early-finished GPUs cannot enter idle mode and waste energy~\cite{pytorch2023ddp}.

To reproduce this energy waste, we need an uneven workload. This is in fact common in post-training with variable-length datasets like ShareGPT~\cite{shareGPT} and UltraChat~\cite{ultrachat}. To avoid wasted computation on padding, variable-length sequences are typically packed into fixed-length batches. Since length distributions differ across data-parallel shards, each GPU ends up with a different number of batches.

We fine-tune a single-layer GPT model with data parallelism on two H200 GPUs over ShareGPT. Due to the natural length distribution of ShareGPT, the two ranks end up with a 33\% difference in batch counts. To quantify the energy waste, we manually implement an early-exit protocol in which a finished GPU notifies the remaining GPUs and the training job is reconfigured to continue with a reduced set of participating GPUs. Thus, the finished GPU can go to the idle mode quickly. Figure~\ref{fig:study_redundant_energy} compares the power traces of \textit{Join} and our early-exit protocol. Both methods have the same training time and model accuracy, but early-exit lets the faster GPU enter idle mode, reducing total energy consumption by about 23\%.

Because \textit{Join} only adds communication on the already-finished GPUs, it does not affect end-to-end training time or model accuracy. As a result, existing performance profilers cannot flag this behavior as anomalous. Even if an energy profiler reports high power usage on the finished GPUs, it is difficult to attribute the root cause to the all-reduces inside \textit{Join} that prevent the GPU from entering idle mode.

\begin{figure}[t]
    \centering
    \includegraphics[width=0.9\columnwidth]{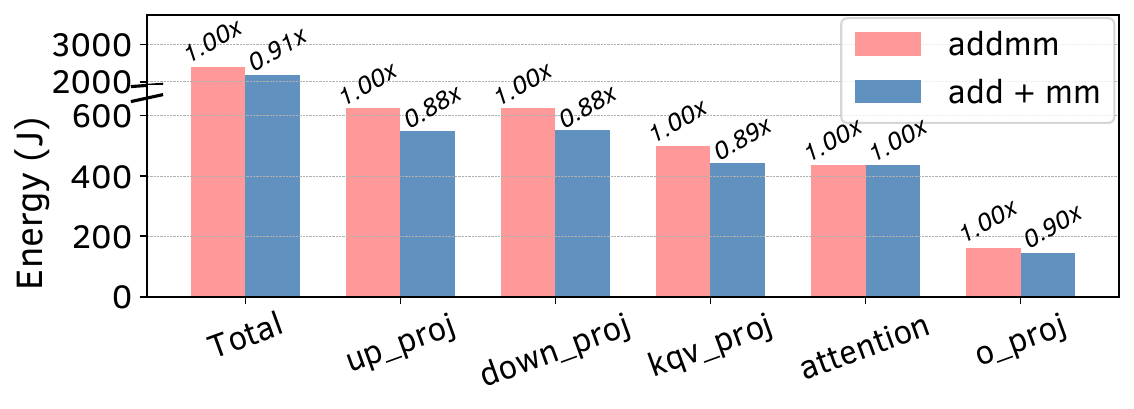}
    \caption{Total energy consumption and breakdowns of top 5 operators in HuggingFace Transformers.}
    \label{fig:study_api_energy}
\end{figure}

\textbf{Case 2: High Energy Operator.} HuggingFace Transformers has a custom \texttt{Conv1D} layer that invokes \texttt{torch.addmm} to fuse the bias addition into the GEMM kernel. In contrast, PyTorch's \texttt{nn.Linear} layer invokes \texttt{torch.matmul} followed by a separate bias add. Intuitively, the fused \texttt{torch.addmm} should be more efficient. However, developers find that on large FP16 shapes, cuBLAS selects a less efficient GEMM kernel for \texttt{torch.addmm} than for \texttt{torch.matmul}, making the fused path have more global memory reads and become more energy-hungry than the unfused one~\cite{torch-141210}.

To reproduce the energy issue, we ran a single-layer GPT-2 model with a batch size of 8 and an input length of 1024 on an NVIDIA H200 GPU. As shown in Figure~\ref{fig:study_api_energy}, the \texttt{torch.addmm} consumes 10.0\% more energy than \texttt{torch.matmul} and \texttt{torch.add} combined.
On the other hand, the inference time for both is the same. This might be because the workload is compute-bound and thus the memory usage is not in the critical path.
This software energy waste is therefore difficult to detect using traditional performance profiling. 

\textbf{Case 3: Configuration Mismatch.} Stable Diffusion~\cite{stable-diffusion} is a widely deployed image generation model. It uses PyTorch as its underlying framework, which uses higher-performance TensorFloat-32 (TF32) matrix multiplication as the default configuration before v1.11. However, PyTorch disables TF32 for \texttt{matmul} by default since v1.12 due to numerical-precision concerns for general-purpose users~\cite{Pytorch_tf32}. Stable Diffusion's default configuration does not override this setting, even though its diffusion-based generative workload is robust to the small precision loss from TF32. As a result, it incurs both performance loss and unnecessary energy consumption, increasing the end-to-end energy consumption by 12.5\%~\cite{sd-forge-279}.

\subsection{Approach to Address Energy Inefficiency}

To detect energy inefficiency, an intuitive way is to use traditional performance profilers, which primarily trace execution time and resource utilization. 
However, these tools are fundamentally insufficient for energy analysis because execution time may not be a perfect proxy for energy consumption.
For example, in Case 2 (\S\ref{sec:motivation_study}), a performance profiler would report no significant performance degradation since both APIs show similar execution times even if they have completely different energy consumption. More generally, operations off the critical path (most notably asynchronous collective communication, which in distributed training is overlapped with computation by design)
do not affect end-to-end performance and therefore remain invisible to performance profilers, even when they consume large amounts of energy.

Since performance profilers are insufficient for detecting and diagnosing energy inefficiency, we need accurate energy profiling tools to guide the optimization. However, existing energy profilers~\cite{Zeus2023NSDI,Batchsizer2021ASPDAC} suffer from two key limitations. First, their sampling rates are too low to provide fine-grained measurements, which is needed to capture the energy consumption of short-lived operators with milliseconds of latency. Second, and more fundamentally, even with operator-level energy data, it remains unclear whether an energy hotspot truly represents an optimization opportunity. For example, matrix multiplication is almost always an energy hotspot, but its implementation may already be optimally efficient. Thus, operator-level profiling alone cannot determine whether excess energy use is avoidable.

Another potential solution to pinpointing software energy waste is to identify code patterns associated with waste using static analysis. However, as shown in \S\ref{sec:motivation_study}, energy waste has diverse root causes, and diagnosing them requires deep semantic and domain-specific knowledge that static analyzers lack. For example, in case 3, locating an incorrectly set \texttt{allow\_tf32} flag as the root cause among thousands of PyTorch configurations requires domain knowledge about NVIDIA GPUs, making static analysis alone insufficient.

These limitations reveal that it is challenging to address the energy waste without knowing what a more efficient implementation looks like.

\begin{figure}[t]
  \centering
  \includegraphics[width=0.98\columnwidth]{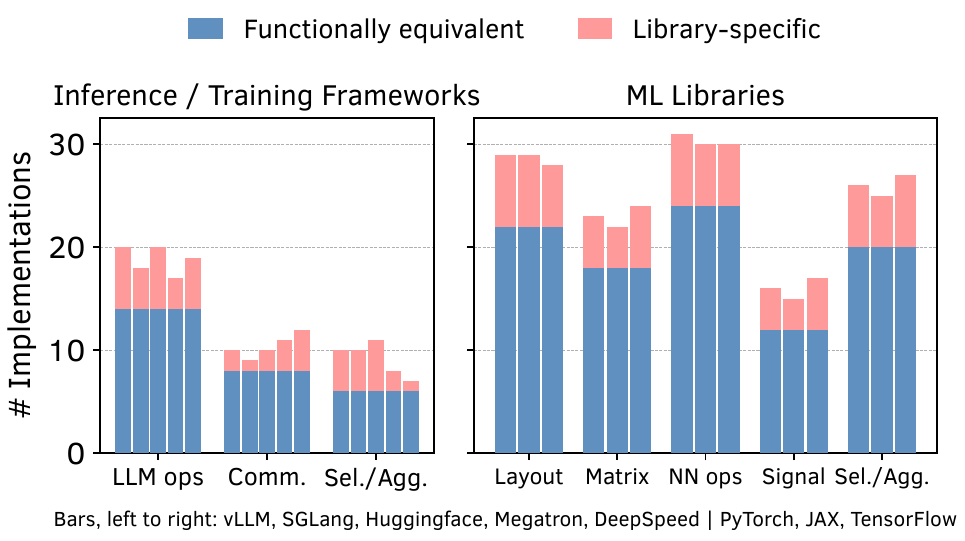}
  \caption{Number of functionally equivalent implementations per category across top ML frameworks and libraries.}
  \label{fig:similarity}
\end{figure}

\begin{figure}[t]
  \centering
  \includegraphics[width=0.75\columnwidth]{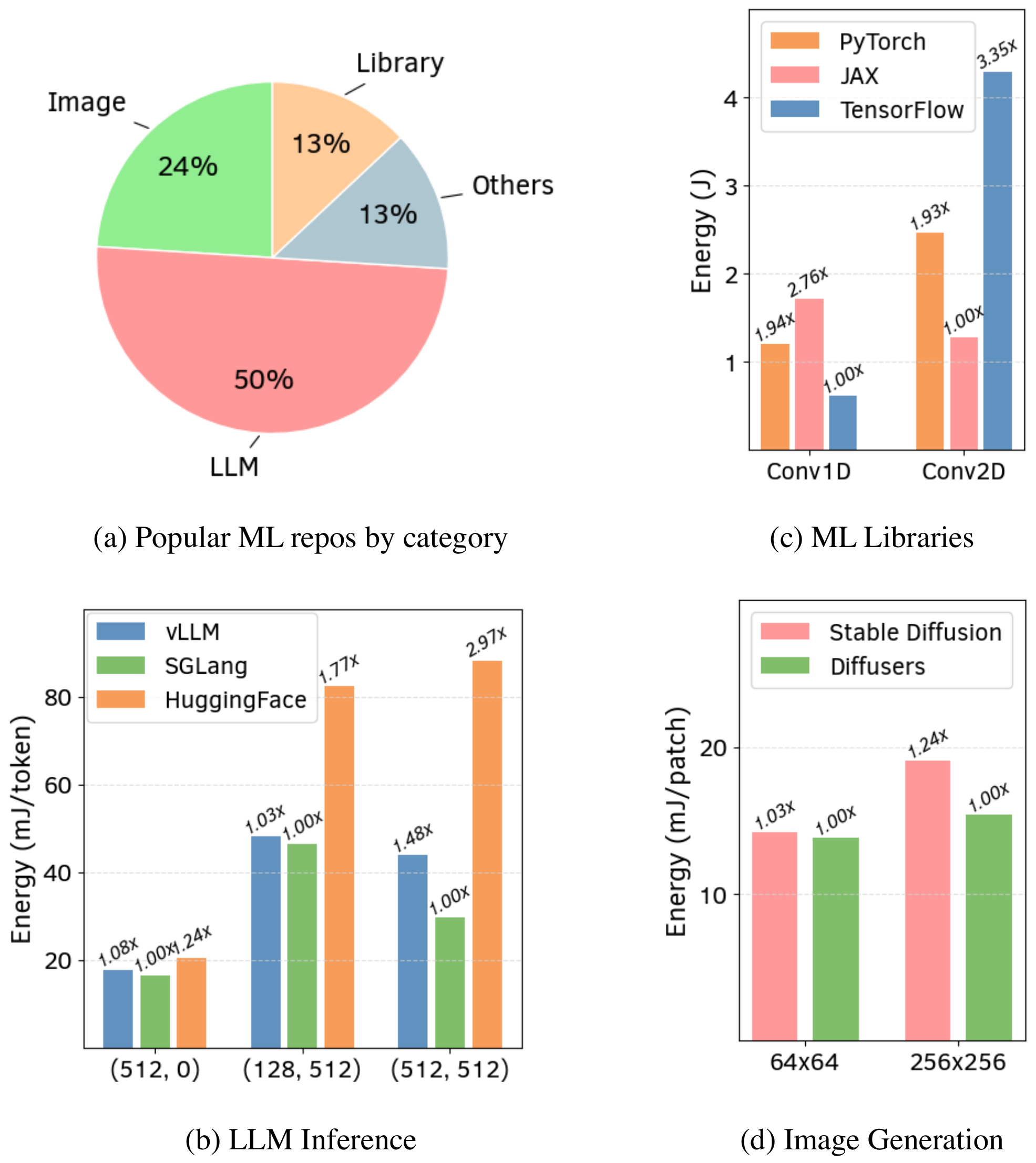}
  \caption{(a) Summary of popular machine learning repositories grouped by category. (b) Energy consumption of LLM inference systems; $(x, y)$ means each request contains $x$ input and $y$ output tokens. (c) Energy consumption of operators in different ML libraries. (d) Energy consumption of different image generation systems.}
  \label{fig:insight}
  \vspace{-1em}
\end{figure}

\section{Key Insights}

\begin{figure*}[t]
    \centering
    \includegraphics[width=0.95\textwidth]{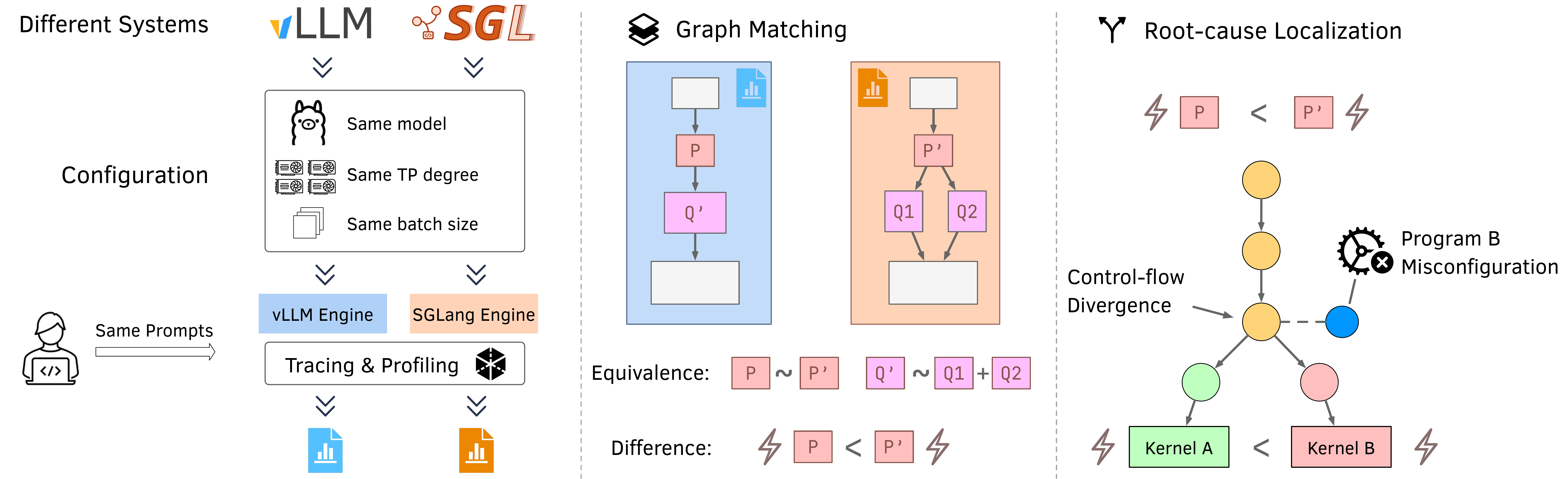}
    \caption{System architecture and workflow of \sysname{}.}
    \label{fig:overview}
\end{figure*}

In this section, we highlight our key insights that motivate the design of \sysname{}.

\subsection{Similarity across ML Systems}

In recent years, generative AI technologies have made ML a vibrant and competitive ecosystem where multiple systems often solve the same problem.
At the library level, PyTorch, TensorFlow, and JAX offer competing implementations for core deep learning functionalities.
For specific tasks like high-throughput LLM inference, frameworks such as vLLM and SGLang provide similar solutions with different custom implementations.
While their internal logic and code structure may differ, these systems do ultimately perform the same underlying mathematical computations to produce correct results.
Figure~\ref{fig:similarity} shows that every category among the most popular ML repositories contains multiple functionally equivalent implementations.

\subsection{Different Energy Consumption}\label{sec:insight-comparison}

The existence of these functional alternatives motivates a question: do different implementations exhibit different energy profiles when performing the same task? To investigate this, we conducted an empirical study comparing state-of-the-art systems across three representative categories:
\begin{itemize}
    \item LLM inference frameworks: we compare the energy consumption (in Joules per token) of vLLM~\cite{vllm}, SGLang~\cite{sglang}, and HuggingFace Transformers~\cite{hf-transformers}. Each system serves a Llama-3-8B model with an identical scheduling policy and workload.
    \item ML libraries: we benchmark the energy consumption of the convolution API in  PyTorch~\cite{pytorch}, TensorFlow~\cite{tensorflow}, and JAX~\cite{jax} using identical input tensors (batch size 128, hidden dimension 512).
    \item Image generation frameworks: We measure and compare the total energy consumption of HuggingFace diffusers~\cite{diffusers} and stable diffusion~\cite{stable-diffusion} to generate an image with the stable-diffusion-3 model.
\end{itemize}

The comparison result is shown in Figure~\ref{fig:insight}.
Surprisingly, we find that they have a significant difference in energy consumption when doing the exact same task.
For example, the difference in the end-to-end energy consumption of SGLang and HuggingFace Transformers can be as high as 2.97$\times$.
Even if we just compare a single operator, the energy difference of the 2D convolution operator between JAX and TensorFlow can also reach 3.35$\times$.

\subsection{Differential Energy Debugging}

Building on these observations, we propose differential energy debugging to leverage the widely existing similarity across different ML systems to detect and diagnose software energy waste. 
The fundamental principle is straightforward: when different systems are doing the same task, their energy consumption should theoretically be comparable, and significant deviations indicate potential inefficiencies. 
However, realizing this approach encounters three principal challenges:
(1) At what abstraction level should we compare different ML systems to effectively detect software energy waste? (2) How do we perform comparisons across systems that implement similar functionality with vastly different approaches?
(3) How do we diagnose the root cause of energy waste when the detected symptoms may not directly reveal the underlying inefficiencies?

\section{Design of \sysname{}}\label{sec:design}

To address these challenges, we design and implement \sysname{}, an end-to-end differential energy profiler that systematically detects and diagnoses software energy waste in ML systems. Figure \ref{fig:overview} presents the system architecture and workflow of \sysname{}.

\sysname{} realizes our differential energy debugging through three steps: (1) it uses operators as the comparison granularity to balance detection precision with analysis feasibility (\S\ref{sec:design-diffing});
(2) it identifies semantic equivalence between operators across different systems to locate the inefficient operators (\S\ref{sec:design-detect});
(3) it diagnoses the root cause by static analysis (\S\ref{sec:design-diagnose}), supported by fine-grained energy tracing and profiling capabilities (\S\ref{sec:impl}).

The usage of \sysname{} is straightforward: developers simply need to identify two systems with similar functionality, provide identical workloads to both systems, and \sysname{} automatically detects energy waste and diagnoses its root cause.
The peer need not be a different system: \sysname{} also supports intra-system comparison of configurations, backends, or versions of one system (e.g., PyTorch v2.6 vs.\ v2.8), or only the subgraphs two systems share (e.g., the ResNet encoder of a detection and a segmentation model).

\subsection{Operator as the Diffing Granularity}\label{sec:design-diffing}

The first design challenge of differential energy debugging is determining the appropriate abstraction level for comparing energy consumption across different ML systems. This choice fundamentally determines both the detectability of energy waste and whether the findings can be acted on.

Both coarse-grained and fine-grained approaches have significant limitations. Coarse-grained comparisons (e.g., model-level or iteration-level) aggregate energy differences across many operations, making it useless for locating fine-grained root causes. For example, in \S\ref{sec:insight-comparison} we see that HuggingFace Transformers can consume 2.97$\times$ more energy than SGLang when doing the same task, but this data alone does not provide much insight into energy optimization.
Fine-grained approaches (e.g., intra-CUDA kernel analysis) face the opposite problem: excessive implementation diversity creates noise rather than insight, as there can be completely different lower-level implementations.
More critically, fine-grained analysis encounters insurmountable practical barriers that we cannot obtain energy consumption data for individual lines of GPU kernel code, nor can we access the source code of proprietary GPU libraries like cuBLAS.

We identify operators as the right granularity for differential energy debugging.
Operators, such as matrix multiplication (GEMM), convolution, and communication primitives, are fundamental units of an ML model.
They provide semantics since different ML systems performing the same high-level function typically execute effectively the same set of operators.
They offer sufficient granularity for detection, with operator-level energy consumption varying by 1.2$\times$ to 3.8$\times$ across systems in our analysis (\S\ref{sec:insight-comparison}).
Most importantly, operator-level inefficiencies map directly to actionable optimization targets such as algorithm selection or configuration adjustments.

\subsection{Locating Inefficient Operators}\label{sec:design-detect}

Comparing different systems at the operator granularity requires identifying semantic equivalence to determine whether two \textbf{sets of operators} perform the same task.
Traditional approaches like source code analysis are infeasible due to the scale of modern ML systems and the diversity of their implementations.
Semantically equivalent operators often differ in naming and internal logic; for instance, a matrix multiplication might be implemented as \texttt{Linear} in vLLM but as \texttt{Conv1D} (with kernel size 1) in HuggingFace Transformers (Figure~\ref{fig:study_api_energy}).
Such mathematical equivalences are difficult to detect via static analysis.

To overcome these barriers, instead of the source code, we leverage the \textit{computational graph} (DAG).
In this representation, operators are nodes and tensors are edges.
This structures the problem of matching equivalent computation or data-transfer tasks as \textbf{finding equivalent subgraphs}, allowing us to identify equivalence through data flow patterns rather than syntax.

However, simply checking if two subgraphs produce identical outputs for identical inputs is insufficient, as this ignores crucial context.
For example, query and key projections in an attention module are mathematically identical operations, but they serve fundamentally different roles in the model architecture.
To address this limitation, we develop a more robust approach that leverages the deterministic data flow properties of computational graphs.
Since computational graphs form DAGs, the data flow patterns within them are unique and deterministic for any given model input.
This leads us to define a context-aware criterion for semantic equivalence:

\begin{hypothesis}
Subgraph $G_1$ in application A and subgraph $G_2$ in application B are
\textbf{semantically equivalent} if they are consistently invoked with
\textbf{semantically equivalent} inputs and produce \textbf{semantically
equivalent} outputs.
\end{hypothesis}

\begin{figure}[t]
  \centering
  \includegraphics[width=0.95\columnwidth]{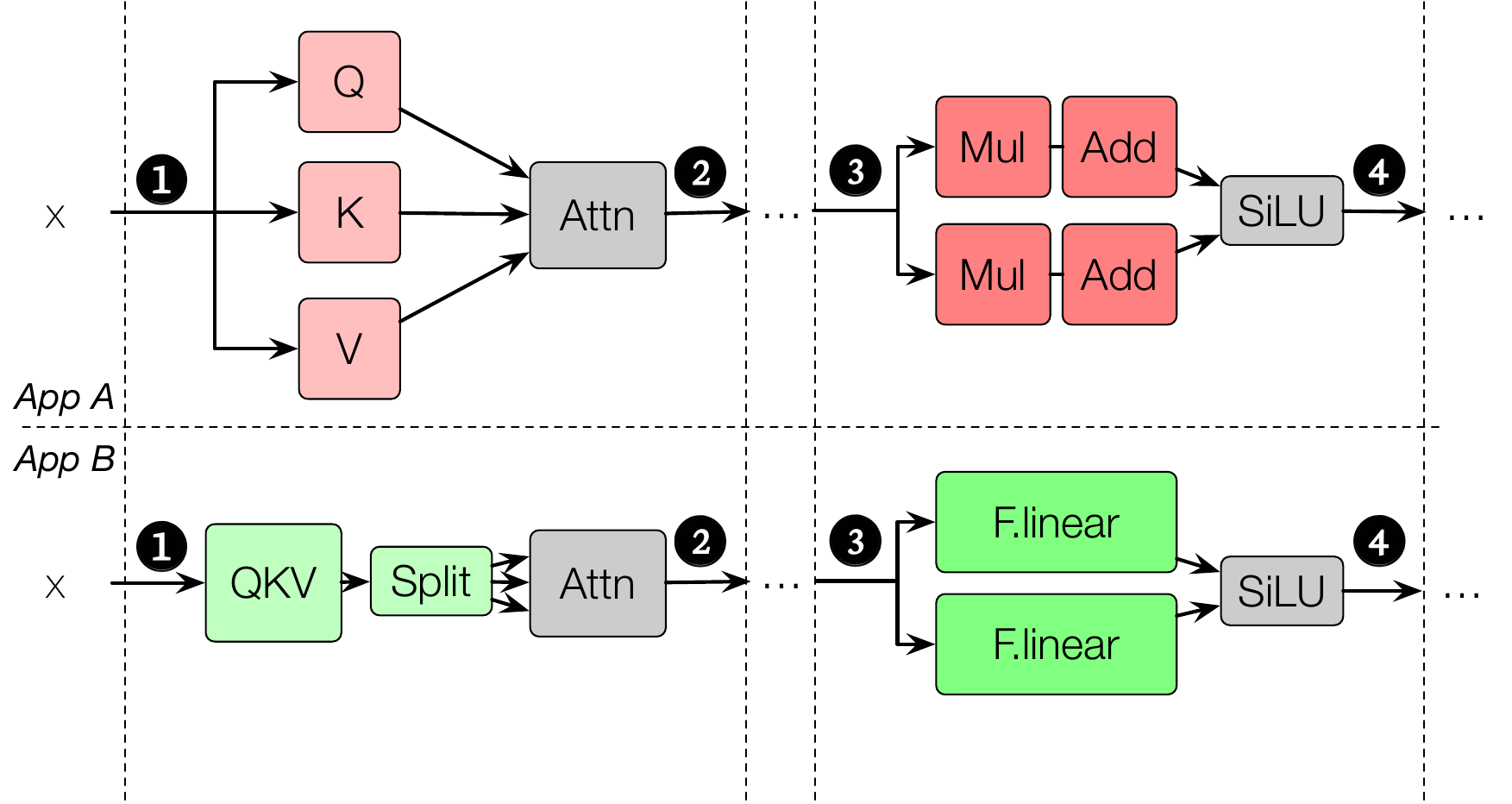}
  \caption{Matching semantically equivalent subgraphs in the computational graph.}
  \label{fig:comp-graph}
\end{figure}

For example, in Figure~\ref{fig:comp-graph}, the pink subgraph in application A (up) formed by \texttt{Q}, \texttt{K}, and \texttt{V} is semantically equivalent to the light green subgraph in application B (down) formed by \texttt{QKV} and \texttt{Split}; the red subgraph formed by \texttt{Mul} and \texttt{Add} is semantically equivalent to the green subgraph formed by \texttt{F.linear}. 
Based on this hypothesis, we first build a tool to match semantically equivalent tensors, then apply it to different systems and computational graphs to find semantically equivalent subgraphs.
In practice, \sysname{} verifies this condition over a
finite set of sampled model inputs (\S\ref{sec:eval-setup}) rather than over all possible inputs.

\noindent\textbf{Matching Equivalent Tensors.}
A naive approach to check tensor equivalence would perform element-wise comparison (e.g., \texttt{torch.allclose}), as commonly done in unit tests.
However, this fails as semantically equivalent tensors can have different layouts across different ML systems.
For example, HuggingFace Transformers and SGLang use different attention backends with distinct tensor layouts: HuggingFace uses \texttt{HND} (attention head first) while SGLang uses \texttt{NHD} (sequence length first). 
These layouts differ only by a \texttt{permute} operation, but this breaks straightforward element-wise matching.

To address this challenge, we propose a robust matching method based on the statistical properties of tensors.
We observe that layout transformations (e.g., \texttt{permute}, \texttt{reshape}) merely reorder entries without changing the singular-value spectra of tensor unfoldings.
For an $r$-way tensor $\mathcal{T}$, we enumerate non-trivial subsets $G \subset[r-1]$ and matricize the tensor by grouping dimensions in $G$ as rows and the complement $G^c$ as columns. This yields $\mathbf{T}{(G)} \in \mathbb{R}^{M_G \times N_G}$ where $M_G = \prod{i \in G} n_i$ and $N_G = \prod_{j \in G^c} n_j$.
We then compute
\begin{align*}
\mathbf{T}_{(G)} = \mathbf{U}_G \boldsymbol{\Sigma}_G \mathbf{V}_G^T, \text{where }  \boldsymbol{\Sigma}G = \text{diag}(\sigma_1, \sigma_2, \ldots, \sigma{\min(M_G, N_G)})
\end{align*}
Here $\sigma(\mathbf{T}{(G)}) = [\sigma_1, \sigma_2, \ldots, \sigma{\min(M_G, N_G)}]^T$ are sorted in descending order.
Then we can compute the multi-mode SVD invariant set:
\begin{align*}
\mathcal{S}(\mathcal{T}) = {\sigma(\mathbf{T}_{(G)}) : G \subset {0,\ldots,r-1}, G \neq \emptyset, G \neq {0,\ldots,r-1}}
\end{align*}
Two tensors are declared equivalent if their invariant sets match under a chosen numerical tolerance $\epsilon$.
\sysname{} compares the unfoldings two tensors share, as a reshape merges modes. The invariant is necessary but not sufficient for layout equivalence, as permuting the indices within one mode preserves it, so a tensor match serves only as a cut point that must also agree with the graph topology (Algorithm~\ref{algo:equiv-graph}).
Despite requiring $2^r-2$ unfoldings, typical tensors have low order ($r \leq 5$), making the overall complexity $O\bigl((2^r-2) \times \max_{G}(M_G^2N_G)\bigr)$ tractable and amenable to parallel SVD implementations.

\begin{algorithm}[t]
\caption{Topology-Aware Subgraph Matching}
\label{algo:equiv-graph}
\begin{algorithmic}
\State \textbf{Inputs}: $G_1, G_2$: computation graphs, $Eq$: pairs of equivalent tensors
\State \textbf{Note:} $n_1 \succ n_2$ means $n_1$ dominates $n_2$ in the graph
\vspace{0.1cm}
\Function{RecursiveMatch}{$G_1, G_2, Eq$}
\State $\mathcal{M} \leftarrow \emptyset$
\Comment{Set of matched subgraphs}
\State $T_1, T_2 \leftarrow$ DominatorTree$(G_1, G_2)$
\State $s_1, e_1 \leftarrow$ Source$(T_1)$, Sink$(T_1)$
\State $s_2, e_2 \leftarrow$ Source$(T_2)$, Sink$(T_2)$
\Comment{Get dominator path}
\State $P_1 \leftarrow \langle t_{1,i} \mid t_{1,1}=s_1, t_{1,n_1}=e_1, t_{1,i} \succ t_{1,i+1} \rangle$
\State $P_2 \leftarrow \langle t_{2,j} \mid t_{2,1}=s_2, t_{2,n_2}=e_2, t_{2,j} \succ t_{2,j+1} \rangle$
\State $\mathcal{E} \leftarrow \{(t_{1,i}, t_{2,j}) \in P_1 \times P_2 \mid (out(t_{1,i}), out(t_{2,j})) \in Eq\}$ 
\State\Comment{Equivalent pairs}

\If{$|\mathcal{E}| = 1$}
    \State \textbf{return} $G_1, G_2$
\EndIf

\For{$k = 1$ \textbf{to} $|\mathcal{E}|$} 
\Comment{Divide and conquer}
    \State $V_{1,k} \leftarrow \{v \in G_1 \mid t_{1,i_k} \succ v, v \succ t_{1,i_{k+1}}\}$
    \State $V_{2,k} \leftarrow \{v \in G_2 \mid t_{2,j_k} \succ v, v \succ t_{2,j_{k+1}}\}$
    \State $G_{1,k} \leftarrow G_1[V_{1,k}]$, $G_{2,k} \leftarrow G_2[V_{2,k}]$
    \State\Comment{$G_{1,k}$ and $G_{2,k}$ are equivalent}
    
    \State $\mathcal{M}_{sub} \leftarrow$ RecursiveMatch$(G_{1,k}, G_{2,k})$
    \State $\mathcal{M} \leftarrow \mathcal{M} \cup \mathcal{M}_{sub}$
\EndFor

\State \textbf{return} $\mathcal{M}$
\EndFunction
\end{algorithmic}
\end{algorithm}

\noindent\textbf{Matching Equivalent Subgraphs.}
However, even with this efficient tensor matching algorithm,
it is still non-trivial to apply the hypothesis to a large computational graph.
A naive approach that checks all possible subgraph pairs between graphs $G_1$ and $G_2$ requires $O(2^{|G_1|} \cdot 2^{|G_2|})$ comparisons, which is unacceptable for large graphs with thousands of operators.

To address this challenge, we design a topology-aware approach that exploits the structure of computational graphs to prune the search space.
Our key insight is that if two graphs are semantically equivalent, their dominator trees also exhibit similar patterns.
This allows us to use a divide-and-conquer strategy to recursively match subgraphs.

Algorithm~\ref{algo:equiv-graph} presents our approach. 
We first perform pairwise tensor matching across all edges in $O(|G_1| \cdot |G_2|)$ time to identify semantically equivalent tensor pairs.
Then, for two graphs with the same input and outputs (which, by hypothesis, are semantically equivalent), we construct their dominator trees $T_1$ and $T_2$ and extract the dominator paths $P_1$ and $P_2$ from source to sink nodes.
We identify equivalent tensor pairs $\mathcal{E}$ along these dominator paths and use them as "cut points" to recursively divide the graphs into smaller subgraphs until there are no finer subgraphs.

Figure~\ref{fig:comp-graph} illustrates this process:
by tensor matching, we identify semantically equivalent tensors marked with the same labels (1-4) across both applications.
These equivalent tensor pairs serve as natural cut points (shown by dashed vertical lines) that divide the computational graphs into manageable subgraphs.
For instance, App A's attention mechanism (Q, K, V, and Attn) is matched against App B's equivalent subgraph (QKV, Split, and Attn), while the subsequent feed-forward components (Mul, Add operations vs. F.linear) form separate comparable subgraphs. This recursive partitioning enables efficient comparison of complex ML computational graphs.

Since all the graphs processed at the same recursion depth share no common edges,
and they are smaller than the graphs processed at the previous depth, the recursion depth and the total complexity of each depth are all $O(N)$,
where $N=\max(|G_1|, |G_2|)$.
Therefore, the overall complexity of the subgraph matching is bound to $O(N^2)$.

\begin{algorithm}[t]
\caption{Finding Deviation Point and Key Variable}
\label{algo:root-cause-localization}
\begin{algorithmic}
\State \textbf{Inputs}: $path_1, path_2$: backtraces of two invoked GPU kernels; $k_1, k_2$: GPU kernels invoked by the applications
\Function{FindDeviationPoint}{$path_1, path_2, k_1, k_2$}
\For{$i = 0$ to $\min(|path_1|, |path_2|) - 1$}
    \If{$path_1[i] \neq path_2[i]$}
        \State $func \leftarrow path_1[i-1]$ 
        \State\Comment{Last common function before deviation}
        \State \Return \textsc{FindKeyVar}($func$)
    \EndIf
\EndFor
\EndFunction

\Function{FindKeyVar}{$func$}
    \State Instrument() 
    \Comment{Instrument all basic blocks for tracing}
    
    \State $trace_1, trace_2 \leftarrow$\textsc{RunAndCollect}()
    \State\Comment{Re-run with instrumentation to get block traces}
    
    \For{$i = 0$ to $\min(|trace_1|, |trace_2|) - 1$}
        \If{$trace_1[i] \neq trace_2[i]$}
            \State $lastCommon \leftarrow trace_1[i-1]$
            \State $control \leftarrow$ GetControlInstruction($lastCommon$)
            \State $key \leftarrow$ ExtractControlVariable($control$)
            \State\Comment{Handle branch, switch, or function pointer}
            \State \Return $key$
        \EndIf
    \EndFor
\EndFunction
\end{algorithmic}
\end{algorithm}

\subsection{Root-cause Diagnosis}\label{sec:design-diagnose}

After identifying the subgraphs with higher energy consumption but no performance or accuracy gain compared to their semantically equivalent alternatives, we diagnose the root cause of the inefficiency.
When applications use different API combinations to implement the same functionality, the diagnosis is straightforward---developers can directly replace the inefficient combination with the more efficient alternative.

The primary challenge arises when both applications call the same library API but different configurations cause the underlying library to select different implementations. For example, in case 3 in \S\ref{sec:motivation_study}, both versions call the same API but invoke different CUDA kernels due to different values of the global flag 
\texttt{torch.}\hspace{0em}\texttt{backends.}\hspace{0em}\texttt{matmul.}\hspace{0em}\texttt{allow\_tf32}
, which affects PyTorch's internal dispatch logic.

To diagnose such cases, we initially considered tracing the complete execution path from the library API entry point (e.g., \texttt{torch.matmul}) down to the actual GPU kernel invocation (e.g., \texttt{cublasSgemm}), inspired by the inflection point hypothesis~\cite{Kairux2019SOSP}.
This hypothesis suggests that in distributed systems, the first point where execution traces diverge between a working and failing system indicates the root cause of the failure.
In our context, this would mean comparing the execution traces of energy-efficient and energy-inefficient implementations to find where they first differ and treating that as the root cause.

However, we found that this approach is inadequate for diagnosing energy waste in ML systems.
In the above example, the first deviation occurs when PyTorch accesses its dispatch table, because the two applications have different internal states.
Yet the actual root cause is much deeper: the \texttt{CUDABlas} function reads the \texttt{allow\_tf32} flag and branches to different code paths, ultimately selecting different CUDA kernels with different energy characteristics.
This example shows that execution traces between different ML applications naturally contain numerous differences in framework internals, CPU buffer allocations, and auxiliary operations, most of which are irrelevant to energy consumption.
This makes us realize that we only need to trace back to understand why GPU kernels consume different amounts of energy—either because different kernels are invoked, or because the same kernel receives different parameters or configurations.

Based on this insight, we develop a targeted approach that focuses specifically on differences between GPU kernels.
The detailed approach is shown in Algorithm~\ref{algo:root-cause-localization}.
Instead of comparing entire execution traces, we extract only the call paths that directly lead to GPU kernel invocations and find deviation points using \textsc{FindDeviationPoint}. 
Then, the \textsc{FindKeyVar} function analyzes these deviation points by instrumenting relevant functions with basic block tracing, re-running both applications, and comparing execution paths to identify the specific branching condition or variable that leads to different GPU kernel selections.
Once we identify the key variable, we perform backward data-flow analysis to trace it to its ultimate source, typically a configuration parameter or API argument that developers can modify to eliminate the energy waste.

\section{Implementation}\label{sec:impl}

To realize the differential debugging approaches in \S\ref{sec:design},
we build a tracing and profiling framework to capture the control-flow, data-flow (\S\ref{sec:impl-software}), and energy consumption (\S\ref{sec:impl-physical}) in ML systems.

\subsection{Tracing Software Events}\label{sec:impl-software}

Tracing ML applications presents unique challenges due to cross-layer Python, C/C++, and GPU execution with asynchronous programming models. We address this through a multi-layer tracing approach that captures GPU events using CUPTI Activity API for kernel executions, memory transfers, and timestamps with minimal overhead.

We intercept ML library calls~\footnote{e.g., PyTorch's \texttt{aten::addGlobalCallback}} and CUDA runtime functions via CUPTI Callback API to capture both high-level operations and low-level GPU invocations.
To establish complete execution context, we capture native C/C++ call stacks using \texttt{libunwind} and Python via \texttt{PyEval\_SetProfile} hooks.
For eager-mode ML frameworks which don't inherently have an exposed computational graph, \sysname{} obtains the computational graph during runtime by executing the program.
Specifically, on PyTorch, it compiles the model callable with TorchDynamo and a custom backend that wraps every FX node.
The backend then records the node's input and output tensors as well as the energy consumption without requiring model rewriting.

In order to correlate asynchronous GPU events with CPU calls, we use CUPTI correlation IDs that uniquely link CPU-side API calls to their corresponding GPU kernel executions, enabling us to construct unified traces reflecting the relationship between CPU scheduling and GPU execution.

\subsection{Energy Profiling}\label{sec:impl-physical}

To support the analysis approaches in \S\ref{sec:design}, \sysname{} needs fine-grained attribution of GPU energy use to individual kernels, which often run in tens to hundreds of microseconds.
Native GPU energy profiling libraries provided by GPU vendors, including NVIDIA management library (NVML) and AMD system management interface, expose GPU power counters but with an update frequency of only 10-50 Hz, and can be delayed by hundreds of milliseconds~\cite{Parttime2024SC}.
This results in a divergence from ground truth by a physical power meter by up to 80\%.
As a result, vendor-provided energy profiling and previous works based on that (e.g., Zeus~\cite{Zeus2023NSDI}) cannot capture the power of a single kernel and provide inaccurate averages across many kernels.
Physical power meters, by contrast, offer microsecond-level resolution and high accuracy, but are not always available to users or easily attached to the server.

\sysname{} incorporates both approaches in a modular design. Users can either use a power meter for direct runtime measurement or use \sysname{}'s replay-based software energy profiling approach.
Specifically, if the hardware power meter is not available, \sysname{} will first use the coarse-grained native GPU energy interface to record iteration-level energy consumption.
If an energy difference is detected across frameworks, we will replay this iteration with recorded input and outputs.
During this process, we replay each operator to have an execution time long enough to average out delays and stabilize readings from GPU energy interfaces, by operator-level framework interception.
This enables accurate power measurement even when users cannot install a physical power meter.

\section{Evaluation}

\begin{table*}[h]
  \centering
  \begin{tabular}{llll}
    \toprule
    \textbf{Id} & \textbf{Case} & \textbf{Category} & \textbf{Description} \\
    \midrule
    c1 & \texttt{vllm-9471} & Misconfiguration & Prefill attention consumes more energy with tensor cores disabled. \\
    c2 & \texttt{vllm-10811} & Redundant & Decode attention incurs energy waste via redundant data copy. \\
    c3 & \texttt{sglang-5128} & API misuse & Top-k implementation launches energy-inefficient APIs. \\
    c4 & \texttt{megatron-543} & Redundant & Redundant \texttt{repeat\_interleave} results in energy waste. \\
    c5 & \texttt{hf-14450} & Misconfiguration & Default tensor format causes energy-intensive layout transformations. \\
    c6 & \texttt{hf-34570} & API misuse & \texttt{torch.linalg.eigvals} selects energy-inefficient kernels. \\
    \midrule
    c7 & \texttt{diffusers-12131} & API misuse & Unnecessary \texttt{concat/split} ops consume extra memory access energy. \\
    c8 & \texttt{sd-279} & Misconfiguration & Linear layers fail to utilize energy-efficient tensor core instructions. \\
    \midrule
    c9 & \texttt{pytorch-181115} & Redundant & \texttt{dist.Join} prevents a finished GPU from going to idle mode. \\
    c10 & \texttt{pytorch-141210} & API misuse & \texttt{torch.addmm} selects kernels with higher energy consumption. \\
    c11 & \texttt{pytorch-28224} & Misconfiguration & Suboptimal flags cause CPU busy-waiting, preventing low-power states. \\
    c12 & \texttt{pytorch-76012} & API misuse & Non-contiguous inputs in \texttt{LayerNorm} trigger inefficient access patterns. \\
    c13 & \texttt{pytorch-141822} & API misuse & \texttt{F.cross\_entropy} launches kernels with higher energy consumption. \\
    c14 & \texttt{jax-28614} & API misuse & \texttt{jax.scipy.signal.stft} calls inefficient low-level APIs. \\
    c15 & \texttt{jax-9239} & Redundant & Redundant computations in \texttt{jax.scipy.linalg.expm}. \\
    c16 & \texttt{tf-60772} & API misuse & \texttt{count\_nonzero} triggers implicit energy-inefficient data copies. \\
    \bottomrule
    \end{tabular}
  \caption{Known issues and their descriptions.}
  \label{tab:known-issues}
\end{table*}

Our evaluation aims to answer the following questions:
\begin{itemize}
  \item How effective is \sysname{} in detecting and diagnosing software energy waste?
  \item Can \sysname{} uncover unknown software energy waste?
  \item What is the accuracy, sensitivity, and overhead of the proposed designs?
\end{itemize}

\subsection{Setup}\label{sec:eval-setup}

\noindent\textbf{Testbed.}
We run the evaluations on two server setups:
(1) Testbed-A: 1 node with 1 NVIDIA RTX 4090 GPU and 2 AMD Threadripper Pro 5955WX CPUs;
(2) Testbed-B: 1 node with 8 NVIDIA H200 GPUs connected through NVLink and 2 AMD EPYC 9534 CPUs.

\noindent\textbf{Target Systems.}
We implement and evaluate \sysname{} on 9 state-of-the-art ML systems,
including 
(1) LLM inference and training systems: SGLang \cite{sglang}, vLLM \cite{vllm},
Megatron-LM \cite{megatron-lm}, and HuggingFace Transformers \cite{hf-transformers};
(2) Image generation systems: Stable Diffusion \cite{stable-diffusion} and Diffusers \cite{diffusers};
(3) ML frameworks: PyTorch \cite{pytorch}, JAX \cite{jax},
and TensorFlow \cite{tensorflow}.

\noindent\textbf{Baselines and Configurations.}
We compare \sysname{} with two state-of-the-art profilers for ML systems and their variants:
(1) Zeus~\cite{Zeus2023NSDI} is an energy monitoring and optimization tool.
(2) PyTorch profiler~\cite{pytorch-profiler} is a widely used performance profiler for PyTorch.
It can obtain the latency of each operator in the model.
(3) Zeus-replay, which we implement on top of Zeus with operator-level replay to address limitations described in \S\ref{sec:impl-physical} and enable accurate power monitoring.
Prior differential analysis works for traditional software~\cite{DiffProf2018OSDI,PeerPressure2004OSDI} are excluded because they cannot capture the asynchronous GPU activities and cross-layer events critical to ML workloads.

Notably, these baselines are not designed for diagnosing the root cause of energy inefficiencies.
We tried to use the basic troubleshooting functionality in PyTorch profiler, but it could not detect any energy waste in our evaluation.
Therefore, for fair comparison, we only use them to monitor the model, profile the operator-level performance and energy statistics, and report the rank of each operator.
The rationale here is that if these tools rank a particular problematic operator as a large energy consumer, that will make it easier for the developer to diagnose the underlying potential software energy waste. 

As no differential energy profiler exists for ML systems, we ask whether these single-system profilers make the problematic operator stand out.
For PyTorch profiler, we use the \texttt{key\_averages()} API to get the latency of each operator of all PyTorch-based systems.
For Zeus and Zeus-replay, we first locate the code of each operator in the model, and then manually instrument the source code by wrapping operators with the \texttt{begin\_window} and \texttt{end\_window} APIs.
For Zeus-replay specifically, we implement a loop to execute each operator 1,000 times with identical inputs, verifying outputs to ensure no side effects are introduced.
However, this instrumentation requires manually locating and modifying the source code for every operator, which is time-consuming and not practical for large end-to-end systems.
Consequently, we limit the evaluation of Zeus-based baselines to cases with not more than 100 operators.

For \sysname{}, we configure the detection threshold for energy inefficiency at a 10\% per-subgraph (not end-to-end) energy difference between the semantically equivalent subgraphs.
\label{sec:eval-threshold}We set it from measurement noise: when the same GPT-2 or Llama-3-8B workload is recorded three times, twice from an idle GPU and once after a three-minute heat soak, and compared through the full pipeline, matched subgraphs of over a millijoule differ by at most 3.2\%.
To strictly classify an inefficiency as software energy waste rather than a performance-energy trade-off or an accuracy-energy trade-off, we enforce a tight tolerance: the efficient variant must not have a performance degradation by more than 1\% or an accuracy loss of more than 1\% (measured by element-wise relative difference in output values).
As subgraph comparison cannot see energy saved elsewhere, we validate each candidate by measuring the end-to-end energy with the alternative applied.

\noindent\textbf{Setup and Usage of \sysname{}.}
The manual effort to integrate \sysname{} into target systems is minimal.
For ML frameworks, users can wrap any defined models or operators with \sysname{}'s energy profiler.
For task-specific ML systems, the integration requires only two steps:
(1) Identify the model callable in the system and wrap it with \sysname{}'s energy profiler;
(2) Add annotations to custom operators to specify the input and output tensors.
The manual effort is minimal for all the target systems, with all added cost being less than 100 lines.

To perform differential analysis, \sysname{} aligns the workloads of distinct systems to ensure they execute the same task.
For high-level LLM/ML frameworks like vLLM and HuggingFace Transformers, alignment is achieved simply by loading the same model checkpoint and supplying identical input prompts.
For ML libraries such as PyTorch, JAX, and TensorFlow, \sysname{} maps the inputs and weights of the compared workload to each library's conventions (e.g., torch.nn to JAX and TensorFlow).
This ensures that all libraries execute the same workload, without requiring developers to port full ML models across libraries.

To further ensure robust comparisons, \sysname{} is configured to handle the inherent stochasticity of ML workloads. 
Instead of bitwise matching, \sysname{} employs a numerical threshold that accommodates minor non-deterministic variations, such as those introduced by dropout or floating-point noise.
For operations that inherently diverge based on random states (e.g., random sampling), \sysname{} relies on consistent random seeds to establish equivalence.
In our setup, we synchronize the seeds at the application level to ensure that.

\begin{table}[t]
  \centering
  \begin{tabular}{lccccc}
    \toprule
    \textbf{Id} & \multicolumn{2}{c}{\textbf{\sysname{}}} & \textbf{PyTorch} & \textbf{Zeus} & \textbf{Zeus-replay} \\
    \cmidrule(lr){2-3} \cmidrule(lr){4-6} & Diag. & Diff. & Rank & Rank & Rank \\
    \midrule
    c1 & \cmark & 12.6\% & 42th & - & - \\
    c2 & \cmark & 1.4\% & >100th & - & - \\
    c3 & \cmark & 2.5\% & 2nd & - & - \\
    c4 & \cmark & 6.7\% & 12th & - & 7th \\
    c5 & \cmark & 58.8\% & 13th & - & 2nd \\
    c6 & \cmark & 29.1\% & 1st & 1st & 1st \\
    c7 & \cmark & 6.1\% & >100th & - & 19th \\
    c8 & \cmark & 12.5\% & >100th & - & - \\
    c9 & \cmark & 7.0\% & 35th & 1st* & 17th \\
    c10 & \cmark & 9.1\% & 14th & - & 1st \\
    c11 & \xmark & <1\% & - & - & - \\
    c12 & \cmark & 16.3\% & 1st & - & 1st \\
    c13 & \cmark & 2.6\% & 5th & - & 6th \\
    c14 & \cmark & 7.7\% & - & - & 2nd \\
    c15 & \cmark & 2.1\% & - & - & 3rd \\
    c16 & \cmark & 27.8\% & - & - & 3rd \\
    \bottomrule
    \end{tabular}
  \caption{\sysname{} detection and diagnosis result. - means baselines are unable to perform fine-grained profiling on this case. * means the profiling result is incorrect.}
  \label{tab:known-issues-diagnosis}
\end{table}

\begin{table}[t]
  \centering
  \setlength{\tabcolsep}{3pt}
  \renewcommand{\arraystretch}{1.2}
  \begin{tabular}{ll>{\raggedright\arraybackslash\hyphenpenalty=10000}p{3.0cm}rc}
    \toprule
    \textbf{Id} & \textbf{Case} & \textbf{Description} & \textbf{Saved} & \textbf{St.} \\
    \midrule
    c17 & \texttt{pytorch-157334} & \texttt{Conv2D} is inefficient in \texttt{NCHW}. & 21.2\% & \cmark \\
    c18 & \texttt{tf-96396} & TF conv.\ is inefficient in \texttt{NHWC}. & 49.5\% & \cmark \\
    c19 & \texttt{pytorch-153195} & Default math mode costs more energy. & 85.5\% & \cmark \\
    c20 & \texttt{jax-29875} & Inefficient cuDNN grouped conv. & 55.4\% & \cmark \\
    c21 & \texttt{hf-38977} & LMHead processes redundant tokens. & 8.7\% & \cmark\cmark \\
    c22 & \texttt{vllm-20174} & Inefficient default prefill attention. & 9.9\% & \xmark \\
    c23 & \texttt{hf-39072} & Inefficient resharding in attention. & 9.0\% & \cmark \\
    c24 & \texttt{hf-39073} & Inefficient default GELU backend. & 16.7\% & \cmark \\
    \bottomrule
  \end{tabular}
  \caption{New issues that \sysname{} identifies with their descriptions. St.: \cmark\cmark~fixed by our merged patch; \cmark~confirmed or acknowledged; \xmark~no response.}
  \label{tab:new-issues}
\end{table}

\subsection{Diagnosing Known Issues}\label{sec:eval-known}

To evaluate the effectiveness of \sysname{} in diagnosing known issues,
we collect 16 real-world cases of software energy waste from the target systems.
We collect them from GitHub issues, developer forums, and blog posts.
Since software energy waste is not well-studied,
some of the cases are collected by searching for related keywords
like \texttt{redundant}, \texttt{inefficient}, \texttt{waste}, etc.
As summarized in Table~\ref{tab:known-issues},
those selected cases span across different ML systems
and exhibit different bug patterns,
making them representative for evaluating the effectiveness of diagnosing energy issues.

For end-to-end ML systems, including categories (1) and (2),
we run the specific version of the target systems
with the same configuration and workload provided by the developers.
Then, we find other systems with the same functionality
and run them with the same settings.
For ML frameworks,
we run the cases with an end-to-end workload that contains the inefficient implementation,
which is collected from the bug report or the model library of the target system like \texttt{torch.nn}.
Then, we compare its energy profiles with
those of an alternative implementation from a different version of the target system or other ML frameworks.
If \sysname{} correctly locates the energy inefficient design,
we mark the case as diagnosed.

In total, \sysname{} successfully diagnoses 15 of the 16 cases.
The results are summarized in Table~\ref{tab:known-issues-diagnosis}.
In those diagnosed cases,
\sysname{} correctly locates the energy inefficient design at the level of
(1) problematic API calls for API misuse cases;
(2) the key configuration for misconfiguration cases;
(3) redundant operations for redundant cases.
For example, in case c1,
\sysname{} points out that vLLM passes \texttt{use\_tensor\_cores=False}
to FlashInfer when serving the Llama 3.1-8B model.
\sysname{} fails to detect c11, as this issue manifests as CPU polling with minimal impact on GPU energy consumption.

We also evaluate the difference between the end-to-end energy consumption on the targeted system before and after the issue is fixed.
The average energy reduction is 13.6\% for the diagnosed cases.
In some cases, although the problematic operator itself is not an energy hotspot, optimizing it can still bring non-negligible energy savings.
For example, in case c7, the \texttt{concat} operator consumes less energy than all dense operators (e.g. \texttt{matmul}) in the model.
However, it occurs in each layer of the model, and removing it can reduce the energy consumption of the model by 6.1\%.

We compare \sysname{}'s capability of diagnosing energy inefficiencies with baselines.
As explained in \S\ref{sec:eval-setup}, we report the rank of the performance and energy consumption of the problematic operator
in the end-to-end workload.
In comparison, among the 12 cases that PyTorch profiler can profile, it only ranks c3, c6, and c12 as the top 3 performance bottlenecks.
Zeus can only accurately get the energy consumption of case c6,
where the problematic kernel runs longer than the minimum measurement window of Zeus (100ms).
For case c9, although Zeus can get the energy consumption of the redundant all-reduce operation,
its ranking is incorrect since it just marks the energy of all other operators as 0.
Zeus-replay ranks 3 cases as the top 1 energy hotspot and 7 cases as the top 5 energy hotspots.
However, although it can locate the energy hot spots for some cases, it does not provide information about the root cause of the energy issues, making it difficult for developers to fix them.

\subsection{Exposing New Issues}

Besides diagnosing known issues,
we also evaluate \sysname{}'s ability to expose unknown energy wastes.
Since all the target systems are popular and widely used,
most of the issues we find,
especially those that also cause performance degradation,
have been reported in the open-source community.
Still, we find 8 new energy issues that can be triggered under certain workloads.
We reported all these issues with the alternative designs to the developers,
and so far, 7 of them have been confirmed.
Table~\ref{tab:new-issues} lists the details of the issues.
Applying each proposed fix reduces the energy of the issue's workload by 8.7--85.5\% (Saved in Table~\ref{tab:new-issues}); our fix for c21 has been merged, and developers invited our patch for c24.

When comparing different LLM serving systems,
we serve GPT-2, Llama 3.1 8B, and Llama 3.1 70B models on them with the same configuration and input,
and then use \sysname{} to collect the energy profiles during runtime.
Although the overall logic of these systems is similar,
we found that they call different CUDA kernels in some operators.
In the GELU operator of the GPT-2 model,
HuggingFace Transformers calls a custom implementation with 5 CUDA kernels,
while vLLM calls a fused kernel that reduces the memory read from HBM to SRAM.
This can reduce the energy consumption of the GELU operator by 77.4\% on average and end-to-end energy consumption by 14--17\% across batch sizes from 1 to 32 and sequence lengths from 128 to 1024.
HuggingFace developers have confirmed this case.

When comparing different ML frameworks, we fuzz popular models and operators with random inputs to detect energy inefficiencies.
By comparing PyTorch and TensorFlow convolution operators, we discover that they call different CUDA kernels with layout-dependent energy trade-offs: TensorFlow's custom kernels are more energy-efficient under \texttt{NCHW} layout, while PyTorch's cuDNN grouped-conv kernels perform better under \texttt{NHWC} layout.
This trade-off is reported to PyTorch developers in c17 and TensorFlow developers in c18, both of which have been confirmed by developers.

\subsection{Semantic Equivalence Matching}\label{sec:eval-equiv}

\begin{figure}[t]
\centering
\includegraphics[width=0.95\columnwidth]{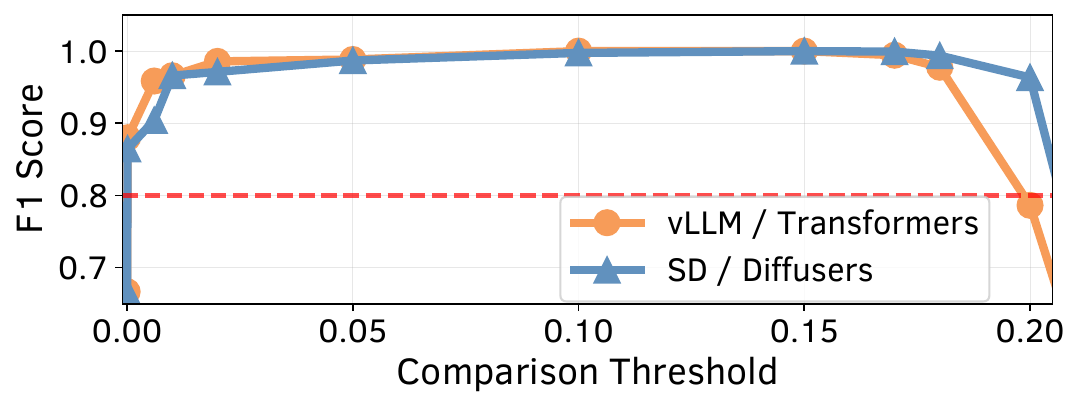}
\caption{The sensitivity of \sysname{}'s semantic equivalence detection algorithm.}
\label{fig:sensitivity}
\end{figure}

\begin{figure}[t]
\centering
\begin{minipage}[t]{0.44\columnwidth}
\centering
\includegraphics[width=0.99\columnwidth]{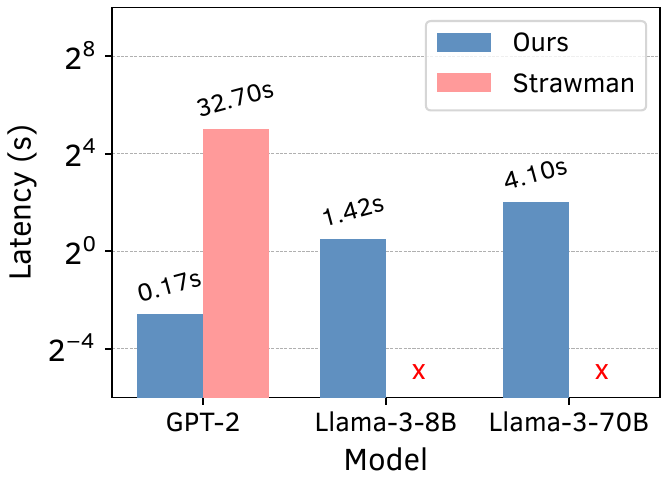}
\vspace{-1em}
\caption{The efficiency of \sysname{}'s semantic equivalence detection algorithm.}
\label{fig:matching-efficiency}
\end{minipage}
\hspace{0.03\columnwidth}
\begin{minipage}[t]{0.44\columnwidth}
\centering
\includegraphics[width=0.99\columnwidth]{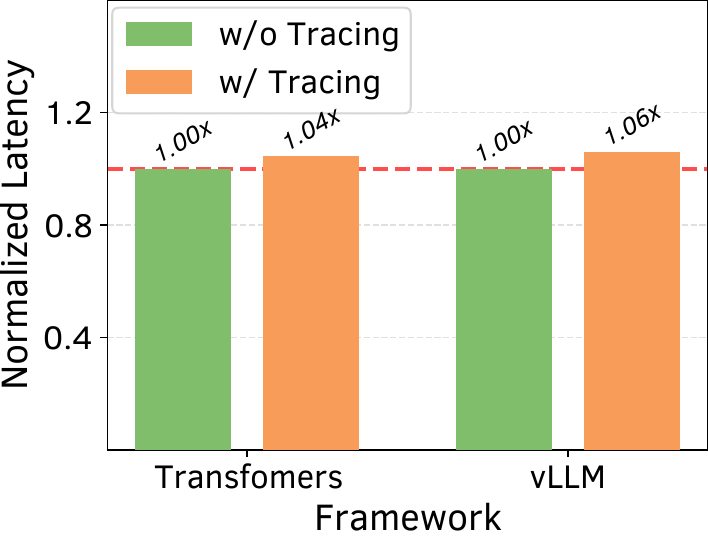}
\vspace{-1em}
\caption{The overhead of \sysname{}'s tracing module on Transformers and vLLM} 
\label{fig:overhead}
\end{minipage}
\end{figure}

\noindent\textbf{Sensitivity.} We first measure the sensitivity of the semantic equivalence matching algorithm to its key parameter: the comparison threshold $\epsilon$.
We vary $\epsilon$ from 1e-7 to 0.2 and evaluate matching accuracy across two workloads: (1) GPT-2 inference (comparing HuggingFace Transformers vs. vLLM) and (2) Stable Diffusion v3 model (comparing Diffusers vs. the reference implementation) using 10,000 identical inputs.
We manually annotated the ground truth for subgraph equivalence to calculate F1 scores.
Figure~\ref{fig:sensitivity} demonstrates that \sysname{} is highly robust to parameter tuning: it maintains an F1 score above 0.8 across a wide magnitude of thresholds ($10^{-4}$ to $1.8 \times 10^{-2}$) and approaches 1.0 within the optimal range.
This confirms that \sysname{} reliably identifies semantic equivalence without requiring manual threshold tuning.

\noindent\textbf{Necessity of the SVD Invariant.}
We compare the invariant against two naive fingerprints, statistics and sorted values, on all 108,889 same-sized pairs of GPT-2 activations from Transformers and vLLM: it reaches an F1 of 1.0 (0.98 and 0.99 for the naive ones), and 0.87 on Llama-3-8B (0.79 and 0.71). The naive fingerprints also accept every copy of an activation that is sorted along one mode or shuffled, of which the invariant rejects 96\% and 100\%.

\noindent\textbf{Efficiency and Scalability.}
We evaluate the efficiency of \sysname{}'s topology-aware matching algorithm against a strawman approach that employs heuristic search with pruning.
Figure~\ref{fig:matching-efficiency} compares the execution latency on computational graphs extracted from vLLM and HuggingFace Transformers when serving GPT-2 and Llama-3-8B.
On the GPT-2 graphs (vLLM: 757 nodes, Transformers: 408 nodes), \sysname{} identifies 71 pairs of equivalent subgraphs (average size: 8.2 nodes; max size: 27 nodes) in just 167ms.
For the larger Llama-3-8B model, \sysname{} demonstrates better efficiency and scalability: while the brute-force approach times out after 5 minute due to combinatorial explosion, \sysname{} completes the matching in 1.4s.
This confirms that our divide-and-conquer strategy effectively prunes the search space, making \sysname{} practical for larger systems and models.
On Llama-3-70B with 4-way tensor parallelism across four H200 GPUs (128-token prompt, 6,206 operators per rank), matching takes 4.1s (Figure~\ref{fig:matching-efficiency}) and isolates the attention subgraph, whose compute operators consume 4--5\% more energy under eager attention than under SDPA, while two SDPA runs differ by at most 0.8\%.

\begin{table}[t]
  \centering
  \setlength{\tabcolsep}{4pt}
  \begin{tabular}{l c cc cc}
    \toprule
    \textbf{Op} & 
    \textbf{Physical} & 
    \multicolumn{2}{c}{\textbf{Zeus}} & 
    \multicolumn{2}{c}{\textbf{\sysname{}}} \\
    \cmidrule(lr){3-4} \cmidrule(lr){5-6}
     & Value & Value & Error\% & Value & Error\% \\
    \midrule
    arange & 266W & 73W & -72.5\% & 255W & -4.1\% \\
    contiguous & 317W & 73W & -77.0\% & 311W & -1.9\% \\
    linear & 455W & 88W & -80.7\% & 459W & +0.9\% \\
    \midrule
    FlashAttn & 695W & 219W & -68.4\% & 693W & -0.3\% \\
    All-Reduce & 365W & 184W & -49.7\% & 363W & -0.6\% \\
    \bottomrule
  \end{tabular}
  \caption{Power measurement methods and relative error rate for different operators.}
  \label{tab:energy-profiling}
\end{table}

\subsection{Tracing Overhead}

We evaluate the runtime overhead introduced by \sysname{}'s software tracing modules by comparing end-to-end latency with and without \sysname{} enabled. Using a mixed workload of 1 prefill request (128 tokens) and 128 decode requests, \sysname{} introduces minimal overhead: 4.4\% for HuggingFace Transformers and 5.9\% for vLLM, as shown in Figure~\ref{fig:overhead}. Additionally, the offline diagnosis module completes within 2 minutes for all evaluated cases. This low overhead demonstrates that \sysname{} can be deployed in production environments without significantly impacting system performance.

\subsection{Energy Profiling}\label{sec:eval-profile}
We evaluate the accuracy of \sysname{}'s energy profiler under software replay mode.
We analyze GPU kernel-level power measurement for energy consumption using three methods on testbed-A with a GPT-2 model (batch size 256, input length 128):
(1) \sysname{} with a physical power meter: we attach an Elmorlabs Power Measurement Device (PMD2)~\cite{pmd2} with an instrumented PCIe riser to measure the ground-truth power draw of the GPU; (2) Zeus~\cite{Zeus2023NSDI}, which uses NVML as its profiling backend; and (3) \sysname{} with operator-level replay.
We evaluate three representative operators in PyTorch: \texttt{arange}, \texttt{contiguous}, and \texttt{linear}.
We also evaluate FlashAttention (causal, batch 8, 1024 tokens) and a 512\,MB All-Reduce across two GPUs on Testbed-B, whose H200 GPUs take no power meter; their ground truth is each operator's power when run alone for 8\,s.
As shown in Table~\ref{tab:energy-profiling},
Zeus exhibits high error rates (\textasciitilde80\%) compared to the ground truth, while \sysname{}'s replay approach achieves accuracy comparable to the physical measurement.
\sysname{} measures these complex kernels within 0.6\%, and across ten captures on each of two H200 GPUs, the power it attributes to an operator varies by at most 1.6\%.
This demonstrates that \sysname{} provides reliable energy consumption measurements in both hardware-accessible and software-only deployment scenarios.

\section{Discussion}\label{sec:discussion}

\noindent\textbf{Limitations.}
\sysname{} makes several assumptions about how it is used.
First, it assumes that a more energy-efficient peer exists, so an inefficiency shared by all peers goes undetected; intra-system comparison (\S\ref{sec:design}) widens the choice of peers.
Second, it assumes that the developer provides inputs that trigger the issue.
The current implementation of \sysname{} also has some limitations, including (1) operator-level visibility: idleness that adds operators, like the zero-valued all-reduces of Case 1, is visible, while CPU busy-waiting (c11) or host-side scheduling delays are not; and (2) hardware dependency: diagnoses that depend on library heuristics (e.g., c10) should be re-validated on the target GPU.

\noindent\textbf{Future Work.}
Differential debugging over dataflow graphs is not specific to energy: comparing latency, memory, or values instead can locate performance, memory, or numerical issues.
We also plan to extend \sysname{} to multi-node deployments and to other accelerators beyond NVIDIA GPUs.

\section{Related Work}

\noindent\textbf{Energy Consumption of ML.} 
Previous work has focused on optimizing ML energy consumption through hardware tuning and model configuration. EnvPipe~\cite{EnvPipe2023atc} and Perseus~\cite{Perseus2024SOSP} leverage DVFS during pipeline bubbles, Zeus~\cite{Zeus2023NSDI} combines adaptive batch sizing with DVFS, and DynamoLLM~\cite{DynamoLLM2025hpca} explores parallelization strategies for inference. While these systems achieve energy savings through model-level optimizations, \sysname{} targets operator-level software inefficiencies.

\noindent\textbf{Differential Analysis} Diffing has been broadly applied across diverse domains. In data systems, data diffing has been used for storage~\cite{rebl2004atc}, computer networks~\cite{PacketCache2008SIGCOMM,SmartRE2009SIGCOMM}, and in-memory data structures~\cite{Laika2008OSDI}, as well as in mobile applications~\cite{Raven2017MobiCom}. Beyond data, execution-trace diffing has been exploited to detect deviations from expected behavior in systems, including failure diagnosis (PeerPressure~\cite{PeerPressure2004OSDI}) and mobile inefficiency (DiffProf~\cite{DiffProf2018OSDI}). \sysname{} extends this line of work by applying differential analysis of program traces with energy profiling to the broader setting of ML frameworks and applications, comparing two executions through graph matching on their dataflow rather than through their configurations or code structure.

Prior work on detecting semantically similar code operates under assumptions that break in ML systems. Approaches like DiffProf~\cite{DiffProf2018OSDI} assume application behavior can be grouped as contiguous blocks of code or call-graph slices, which are treated as coarse-grained tasks for differential energy analysis. In ML systems, however, an operator’s behavior spans Python code, C++ dispatch, GPU libraries, and asynchronous GPU events that cannot be easily grouped into stable code segments via bottom-up analysis. To address this, \sysname{} takes an operator-centric differential analysis. It first partitions the execution using ML operators as the semantic units. It then matches equivalent subgraphs of the two systems' dataflow graphs by the tensors their operators exchange, independent of their code. When \sysname{} detects energy inefficiency in an operator, it then examines the operator’s source code and the control- and data-flow to diagnose how the operator is invoked and how it invokes underlying GPU kernels.

\noindent\textbf{Comparing Divergent Executions.} The idea of leveraging divergent execution to locate root causes has been explored in systems like Kairux~\cite{Kairux2019SOSP}.
Kairux diagnoses failures by identifying the divergent point of the entire control-flow across failing and non-failing executions, isolating the minimal root cause.
Kairux compares two executions of the same program, whose control flows coincide until the failure; two ML systems that implement the same model share no control flow to align, so \sysname{} aligns them by dataflow instead.
In contrast, \sysname{} restricts analysis to the specific invocation path of the inefficient GPU kernel, effectively bypassing the irrelevant framework-level divergences that often precede the actual root cause of energy waste.

\section{Conclusion}
\sysname{} is a differential energy profiler that detects and diagnoses software energy waste by matching equivalent dataflow subgraphs across implementations and comparing their operator-level energy. On nine ML systems, it diagnoses 15 known issues and uncovers 8 new ones, 7 confirmed by developers. As ML training and deployment continue to scale, we believe that tools like \sysname{}, and research on ML energy efficiency more broadly, will see wider adoption.

\section*{Acknowledgments}
We thank our shepherd and the anonymous NSDI reviewers for their helpful comments. We also thank Yibo Wu, Kan Zhu, Hongtao Zhang, Yiyu Liu, and Yingbo Sun, as well as other members and alumni of the UW SyFI Lab, for their insightful discussion during the early stages of this work. This work is supported in part by PRISM, one of the seven centers in JUMP 2.0, a Semiconductor Research Corporation (SRC) program sponsored by DARPA, as well as generous donations from NVIDIA, AMD, Intel, and Arm.

\bibliographystyle{plain}
\bibliography{bib/reference}

\end{document}